\documentclass{article}  %
 \evensidemargin \oddsidemargin

\usepackage{epsfig} %

\usepackage{amsmath} %
\usepackage{amssymb}  %
\usepackage{cite}

\usepackage{color}

\title{\LARGE \bf
A Convex Approach to Steady State Moment Analysis for Stochastic
Chemical Reactions}
\date{}
\author{Yuta Sakurai and Yutaka Hori%
\thanks{This work was supported in part by JSPS KAKENHI Grant Number
JP16H07175, Okawa Foundation Research Grant under grant number 16-10,
Keio Gijuku Academic Development Funds and Research Grant of Keio
Leading-edge Laboratory of Science and Technology. 
\textcopyright 2017 IEEE. Personal use of this material is permitted. Permission from IEEE must be obtained for all other uses, in any current or future media, including reprinting/republishing this material for advertising or promotional purposes, creating new collective works, for resale or redistribution to servers or lists, or reuse of any copyrighted component of this work in other works.
}%
\thanks{Y. Sakurai and Y. Hori are with Department of Applied Physics and Physico-Informatics, Keio University, Japan. 
{\tt\small y.sakurai-5861@keio.jp}, {\tt\small yhori@appi.keio.ac.jp}}%
}
\usepackage{bm}
\def\coloneqq{\mathrel{\mathop:}=}

\begin{document}

\maketitle
\thispagestyle{empty}
\pagestyle{empty}

\begin{abstract}
Model-based prediction of stochastic noise in biomolecular reactions often resorts to approximation with unknown precision.
As a result, unexpected stochastic fluctuation causes a headache for the designers of biomolecular circuits. 
This paper proposes a convex optimization approach to quantifying the steady state moments of molecular copy counts with theoretical rigor.
 We show that the stochastic moments lie in a convex 
 semi-algebraic set specified by linear matrix inequalities.
 Thus, the upper and the lower bounds of some moments  can be computed by a semidefinite program. 
 Using a protein dimerization process as an example, we demonstrate that
 the proposed method can precisely predict the mean and the variance of the
copy number of the monomer protein.

 \end{abstract}

\section{Introduction}

\par
\smallskip
A grand challenge of synthetic biology is to build and control layers of
artificial biomolecular reaction networks that enable complex tasks by
utilizing naturally existing reaction resources in micro-scale organisms
such as E. coli. In control engineering community, many theoretical
tools were developed over the last decade to enable model-guided design
of biomolecular circuits including a fold-change detector
\cite{Guo2014}, oscillators \cite{Elowitz2000,Automatica2011,Automatica2013,Niederholtmeyer2015}, an
event detector \cite{Hsiao2016}, and a light-controlled feedback circuit
for setpoint control
\cite{Milias-Argeitis2011,Milias-Argeitis2016}. Other successfully
implemented examples include logic gates \cite{Moon2012}, a genetic memory
\cite{Yang2014}, and a communication system between cells
\cite{Danino2010}, to name a few. Given the elementary modules of
artificial bimolecular networks, the next step stone is to assemble the circuit parts to build multiple layers of biomolecular networks that can provide practical functions. 

\par
\smallskip
As is the case with mechanical and electrical engineering, guaranteeing
the performance of individual circuit modules at high precision is a key
to successfully building large-scale and robust biomolecular
circuits. Biosystems engineering, however, has a unique challenge, in
contrast with other fields of engineering, that the signal-to-noise
ratio is so low that the heterogeneity of circuit states between
biological cells is almost impossible to avoid. A major source of the
heterogeneity is the stochastic chemical reactions that are caused by
the low copy nature of reactive molecules in a small-volume reactor,
that is, a biological cell. In other words, the events of molecular
collision that fire reactions are more accurately captured by stochastic
process rather than continuous and deterministic process governed by
ordinary differential equations (ODEs). This situation raises a strong
need for the theoretical tools that can rigorously certify the
performance of stochastic chemical reactions using statistical norms
such as mean and variance of molecular copy numbers.

\par
\smallskip
The random fluctuation of the copy number of molecules can be regarded as a Markov process that follows the chemical master equation (CME) \cite{Gillespie1992}. 
Despite a linear ODE, the CME is often hard to solve analytically
because of the infinite dimensionality of the equation. A typical
solution to this problem is to use Monte Carlo simulations
\cite{Gillespie1976}. This approach, however, requires high
computational time. Moreover, a strict error bound is hard to obtain. To
resolve these issues, the finite state project \cite{Munsky2006}
provides a systematic way to truncate the equation with an error bound. Although useful for analyzing transient dynamics, the FSP still suffers from high dimensionality if one wants to evaluate the probability distribution near the steady state, which is often the operation point of interest.

\par
\smallskip
A more direct approach to evaluating the statistics of biomolecular
circuits is to use the moment equation, an infinite dimensional ODE
derived from the CME. Recently developed moment closure approaches
\cite{Zhao2010, Singh2011} allow for approximately solving the ODE by
order reduction technique. The central idea is to approximate the high
order moments in the equation using the lower order moments to derive a
finite order {\it closed} ODE. Using this method, we can efficiently and
directly compute the statistics of the molecular copy numbers of
biomolecular circuits. Although appealing, this approach requires
assuming the underlying probability distribution, which is not {\it
apriori} in practice. As a result, biocircuit designers have to engineer
biomolecular circuits based on approximated statistics with unknown
precision. %

\par
\smallskip
The objective of this paper is to propose an algebraic approach to
rigorously quantifying the steady state statistics of stochastic
biomolecular reactions without assuming underlying probability
distributions. Specifically, we formulate a semidefinite
program that computes the upper and the lower bounds of statistical values of molecular copy
numbers using the moment approach \cite{Lasserre2009}.
During the review process, the authors were informed that 
the same approach has been explored independently by two other research
groups \cite{Ghusinga2016, Kuntz2017}.
In comparison with these works, this paper presents an optimization algorithm that directly computes the upper bound of the 
second order {\it central} moment, or variance, without running multiple
optimizations for computing raw moments, allowing for direct and tighter
quantification of the upper bound.
We illustrate the proposed method using a stochastic protein dimerization process.

\par
\smallskip
The organization of this paper is as follows. In Section II, we
introduce the CME and derive the moment equation. Then, the optimization
problem for moment computation is formulated in Section III. In section
IV, we analyze the mean and the variance of a stochastic dimerization
process. Finally, Section V summarizes our findings and concludes this
paper.

\par
\smallskip
The following notations are used in this paper. 
$\mathbb{N}_0:=\{0,1,2,\cdots\}$.
$\mathbb{N}_+:=\{1,2,3,\cdots\}$.
$\mathbb{R}_+ := \{x \in \mathbb{R}~|~x \ge 0\}$.
${\rm deg}(p(x))$ is the degree of the polynomial $p(x)$. For 
multivariate polynomials $p({\bm x})=\prod_{j=1}^{n}x_j^{p_j}$, ${\rm deg}(p({\bm x})) := \sum_{j=1}^{n}p_j$.

\section{Moment Dynamics of Stochastic Chemical Reactions}
\subsection{Model of stochastic chemical reactions}
We consider a chemical reaction network consisting of $n\in\mathbb{N}_+$
species of molecules and denote the copy number of each molecular
species by
$\bm{x}=[x_1,x_2,\cdots,x_n]^\mathrm{T}\in\mathbb{N}_0^n$. There are $r$
types of reactions in the reaction network, and let $\bm{s}_i=[s_{i1},s_{i2},\cdots,s_{in}]^\mathrm{T}~\in\mathbb{Z}^n$ be the increment of the molecular copy $\bm{x}$ by the $i$-th chemical reaction~($i=1,2,\cdots,r$).

\par
\smallskip
Chemical reactions are caused by collisions of molecules. When the reaction volume is sufficiently large, the copy number of molecules is so large that the collision event occurs almost continuously in time. On the other hand, in small reactor systems such as microbes, the collision events tend to be stochastic as the copy number of molecules is small. This results in the stochastic fluctuation of $\bm{x}$. It is known that stochastic chemical reactions can be modeled by a Markov process. More specifically, we define $P_{\bm{x}}(t)$ as the probability that there are $\bm{x}$ molecules at time $t$. The stochastic chemical reactions can be modeled by the following Chemical~Master~Equation~(CME) \cite{Gillespie1992},
\begin{equation}
\frac{dP\!_{\bm{x}}(t)}{dt}\!=\!\sum_{i=1}^r\!\left\{w_i(\bm{x}-\bm{s}_i)P\!_{\bm{x}-\bm{s}_i}(t)-w_i(\bm{x})P\!_{\bm{x}}(t)\right\}\label{eq:CME},
\end{equation}
where $w_i(\bm{x})$ is the transition rate associated with the $i$-th chemical reaction and is defined by
\begin{equation}
w_i(\bm{x})\!=\!\!\lim_{\Delta t\to 0}\!\!\frac{P_{\bm{x}+\bm{s}_i}\!(t+\Delta t)-P_{\bm{x}}\!(t)}{\Delta t}~~(i=1,2,\cdots,r).\label{eq:wi}
\end{equation}
The transition rate $w_i(\bm{x})$ is characterized by the rate of the
reaction $i~(i=1,2,\cdots,r)$
We assume that all of the $r$ reactions are elementary as non-elementary reactions can always be
decomposed into elementary reactions. For elementary reactions,
the transition rate $w_i({\bm x})$ is a polynomial of $x_i$'s, and we utilize this fact in the following subsection to derive the equations
of stochastic moments.

\par
\smallskip
As ${\bm x}$ is the vector of non-negative positive integers $\mathbb{N}_0^n$, the CME 
(\ref{eq:CME}) is a linear but an infinite dimensional ODE with respect to $P_{\bm
x}(t)$. %
An analytic solution to the CME is, thus, hardly obtained except for some
simple examples. In the next subsection, we derive the equations of
moment dynamics to directly compute the statistical values without computing the distribution $P_{\bm  x}(t)$.

\subsection{The dynamics of moments}
In this section, we derive the model of moment dynamics based on the CME (\ref{eq:CME}). First, we define a stochastic moment by
\begin{equation}
 m_{\bm{\alpha}}\coloneqq\mathbb{E}\left[\prod_{j=1}^nx_j^{\alpha_j}\right]\!=\!\sum_{x_1=0}^\infty\!\sum_{x_2=0}^\infty\!\cdots\!\sum_{x_n=0}^\infty
	\prod_{j=1}^n\!x_j^{\alpha_j}P_{\bm{x}}(t),\label{eq:defmoment}
\end{equation}
where
$\bm{\alpha}\coloneqq[\alpha_1,\alpha_2,\cdots,\alpha_n]\in\mathbb{N}_0^n$.
We refer to the sum of the entries of $\bm{\alpha}$ as the order of
the moment and denote by ${\rm deg}(m_{\bm{\alpha}}) := \sum_{j}\alpha_j$
with a little abuse of notation. 

\begin{figure}[tp]
\centering
\includegraphics[width=6.5cm]{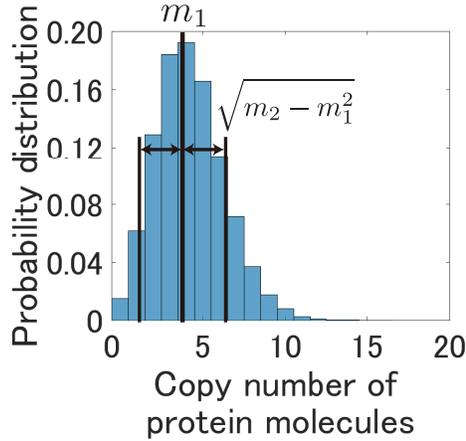}
\caption{An example of probability distribution $P_{\bm x}(t)$ and its relationship with stochastic moments}
\label{fig:conceptfig}
\end{figure}

\par
\smallskip
The stochastic moments carry all the information of the
probability distribution, and some of the low order moments are useful
to define the design specification of stochastic biomolecular circuits.
Figure \ref{fig:conceptfig} illustrates an example of the distribution of a molecular copy number
 $x$ of some biomolecular reaction. %
In this case, the analytic form of the distribution is not known, and thus
it is more appropriate to specify the design goal by statistical
norms such mean $\mathbb{E}[x] = m_1$ and the standard deviation
 $\sqrt{\mathbb{E}[(x-\mathbb{E}[x])^2]} = \sqrt{m_2-m_1^2}$ of the
distribution using the moments.

\par
\smallskip
To derive the dynamics of stochastic moments, we multiply the model
(\ref{eq:CME}) by $\prod_jx_j^{\alpha_j}$ and calculate the expected
value by taking sum over $\mathbb{N}_0^n$ to obtain the following differential equation.
\begin{align}
&\frac{d}{dt}\left\{\sum_{\bm{x}} \left(\prod_{j=1}^nx_j^{\alpha_j} \right)P_{\bm{x}}(t)\right\}\nonumber\\
&\!=\!\sum_{\bm{x}}\prod_{j=1}^nx_j^{\alpha_j}\sum_{i=1}^r\!\left\{w_i(\bm{x}\!-\!\bm{s}_i)P_{\bm{x}-\bm{s}_i}\!(t)\!-\!w_i(\bm{x})P_{\bm{x}}(t)\right\} \label{eq:expectedCME},
\end{align}
where we use $\sum_{\bm{x}}$ to abbreviate the sum over the positive orthant.
\begin{equation}
\sum_{\bm{x}}:=\sum_{x_1=0}^\infty\sum_{x_2=0}^\infty\cdots\sum_{x_n=0}^\infty.
\end{equation}
As $w_i({\bm x})$ is a polynomial of $x_i~(i=1,2,\cdots,n)$, we can
rewrite the equation (\ref{eq:expectedCME}) as
\begin{equation}
\frac{d}{dt}m_{\bm{\alpha}}=\sum_{\beta_1}\sum_{\beta_2}\cdots\sum_{\beta_n}A_{\bm{\alpha},\bm{\beta}}m_{\bm{\beta}}~~~(\bm{\alpha}
 \in \mathbb{N}_0^n), \label{eq:momenteq}
\end{equation}
where $A_{\bm{\alpha},\bm{\beta}}$ are the coefficients of the monomials of the polynomial
\begin{align}
\sum_{i=1}^r&\sum_{\bm{x}} \left\{(x_1+s_{i1})^{\alpha_1}(x_2+s_{i2})^{\alpha_2}\right.\nonumber\\
&\left.\cdots (x_n+s_{in})^{\alpha_n}-x_1^{\alpha_1}x_2^{\alpha_2}\cdots
 x_n^{\alpha_n}\right\}w_i(\bm{x}). \label{eq:momenteqcoefficient}
\end{align}
Note that the range of the summation in (\ref{eq:momenteq}) depends on
the degree of the polynomial (\ref{eq:momenteqcoefficient}). In  many
cases, the higher order moments are necessary to characterize the low
order moments, {\it i.e.,} there are moments $m_{\bm{\beta}}$ satisfying
${\rm deg}(m_{\bm{\beta}})\!>\!{\rm deg}(m_{\bm{\alpha}})$ in the
right-hand side of (\ref{eq:momenteq}).
More specifically, when the reactions are elementary, it is reasonable
to assume that the polynomial order of the transition rates
$w_i(\bm{x})$ are at most two since most practical reactions are
unimolecular or bimolecular reactions \cite{Denisov2003}.
When ${\rm deg}(w_i({\bm x})) = 0$ or $1$, $m_{\bm{\beta}}$ in the right-hand side of equation (\ref{eq:momenteq}) satisfies 
\begin{equation}
 0 \le {\rm deg}(m_{\bm{\beta}}) \le {\rm deg}(m_{\bm{\alpha}}),
	\label{linear-beta-eq}
\end{equation}
implying that the moment equation is closed, {\it i.e}, the dynamics of the $i$-th order moment can be written by
the $i$-th or lower order moments. 
On the other hand, when one or more $w_i({\bm x})$'s are quadratic, {\it
i.e.,} ${\rm deg}(w_i({\bm x})) = 2$, 
\begin{equation}
 0 \le {\rm deg}(m_{\bm{\beta}}) \le {\rm deg}(m_{\bm{\alpha}}) + 1.
		\label{nonlinear-beta-eq}
\end{equation}
Thus, the moment equation (\ref{eq:momenteq}) becomes an infinite
dimensional coupled linear ODE, which is hard to solve analytically.

\section{Moment Analysis of Stochastic Chemical Reactions}
\subsection{Truncation of the moment equation}
When we design stochastic chemical reactions using a genetic circuit,
the specification of the circuit is often given by a set of statistical
constraints such as the mean and the variance of the copy numbers of the
molecules. In this paper, we consider the following problem.

\medskip
\noindent
\textbf{Problem 1.}~For a given CME (\ref{eq:CME}), find the lower and
the upper bounds of the mean and the variance of molecular copy number $\bm{x}$ at steady state.

\medskip
\noindent
To solve this problem, let $dm_{\bm{\alpha}}/dt=0$ in the equation
(\ref{eq:momenteq}), and consider the subset of linear equations by
limiting the equations up to the $\mu$-th order moments. 
Specifically, 
\begin{equation}
0=\sum_{\beta_1}\sum_{\beta_2}\cdots\sum_{\beta_n}A_{\bm{\alpha},\bm{\beta}}m_{\bm{\beta}}~~~\left(\bm{\alpha}\in\mathcal{A}_{\mu}\subset\mathbb{N}_0^n\right),\label{eq:momenteq0}
\end{equation}
where the set $\mathcal{A}_{\mu}$ is defined by
\begin{equation}
 \mathcal{A}_{\mu}\coloneqq\left\{\bm{\alpha}\in\mathbb{N}_0^n~\bigg|~0\le \sum_{i}^{n}\alpha_i\le\mu\right\}.\label{eq:setAmu}
\end{equation}
In what follows, we refer to $\mu$ as the truncation order.
When ${\rm deg}(w_i({\bm x})) = 0$ or $1$, the number of equations is
the same as the number of variables (see (\ref{linear-beta-eq})), and
thus we can compute the unique solution (if the equations are not degenerated).
In the more general case where the transition rates $w_i(\bm{x})$ are quadratic, the
right-hand side of (\ref{eq:momenteq0}) depends on the higher order
moments, {\it i.e.,} ${\rm deg}(m_{\bm{\beta}})= \mu + 1$, as shown in (\ref{nonlinear-beta-eq})
Consequently, the equations become underdetermined, and we can only
conclude that the solution lies on a certain hyperplane unless we
consider an infinite number of equations.

\par
\smallskip
To further specify the solution space of the moments $m_{\bm{\beta}}$,
we utilize the fact that $m_{\bm{\beta}}$ must be the moments of some
(probability) measure defined on the positive
orthant. %
In particular, we use a so-called moment condition to constrain
$m_{\bm{\beta}}$ and formulate a semidefinite program solving Problem
1.

\subsection{Necessary condition for $m_{\bm{\beta}}$ to be moments}
Let $\bm{x}^p$ be the vector that consists of all of the monomial bases
satisfying ${\rm deg}(\prod_jx_j^{p_j})=p$, and $(H)_{p,q}$ be a matrix
of the form
\begin{equation}
(H)_{p,q}\coloneqq\mathbb{E}\left[\bm{x}^p\left(\bm{x}^q\right)^\mathrm{T}\right], \label{eq:hpqmatrix}
\end{equation}
where $\mathbb{E}[\bm{x}^p (\bm{x}^q)^\mathrm{T}]$ represents the
entry-wise expected value of the matrix
$\bm{x}^p(\bm{x}^q)^\mathrm{T}$. In a similar manner, we define
$(H_k)_{p,q}$ by
\begin{equation}
(H_k)_{p,q}\coloneqq\mathbb{E}\left[x_k~\bm{x}^p\left(\bm{x}^q\right)^\mathrm{T}\right]~~~(k=1,2,\cdots,n).\label{eq:hkpqmatrix}
\end{equation}
It should be noted that the entries of the matrices $(H)_{p,q}$ and
$(H_k)_{p,q}$ can be written with $m_{\bm \beta}$ satisfying
\begin{align}
	{\rm deg}(m_{\bm \beta}) = %
	 \begin{cases}
		p + q & {\rm for}~(H)_{p,q}\\
		p + q + 1 & {\rm for}~(H_k)_{p,q}
	 \end{cases}.
\end{align}
Using the matrices $(H)_{p,q}$ and $(H_k)_{p,q}$, we define the
following block Hankel matrices $H^{(\gamma_1)}(\{m_{\bm{\beta}}\})$ and
$H_k^{(\gamma_2)}(\{m_{\bm{\beta}}\})~(k=1,2,\cdots,n)$. 
\begin{equation}
H^{(\gamma_1)}\!(\{m_{\bm{\beta}}\})\!\coloneqq\!\left[
\begin{array}{cccc}
\!\!\!(H)_{0,0}\! & \!(H)_{0,1}\! & \!\cdots\! & \!(H)_{0,\gamma_1}\!\!\!\\
\!\!\!(H)_{1,0}\! & \!(H)_{1,1}\! & \!\cdots\! & \!(H)_{1,\gamma_1}\!\!\!\\
\!\!\!\vdots\! & \!\vdots\! & \!\ddots\! & \!\vdots\!\!\!\\
\!\!\!(H)_{\gamma_1,0}\! & \!(H)_{\gamma_1,1}\! & \!\cdots\! & \!(H)_{\gamma_1,\gamma_1}\!\!\!
\end{array}
\right],\label{eq:momentmatrix}
\end{equation}
\begin{equation}
H_k^{(\gamma_2)}\!(\{m_{\bm{\beta}}\})\!\coloneqq\!\left[
\begin{array}{cccc}
\!\!\!(H_k)_{0,0}\!\! &\!\! (H_k)_{0,1}\!\! &\!\! \cdots\!\! &\!\! (H_k)_{0,\gamma_2}\!\!\!\\
\!\!\!(H_k)_{1,0}\!\! &\!\! (H_k)_{1,1}\!\! &\!\! \cdots\!\! &\!\! (H_k)_{1,\gamma_2}\!\!\!\\
\!\!\!\vdots\!\! &\!\! \vdots\!\! &\!\! \ddots\!\! &\!\! \vdots\!\!\!\\
\!\!\!(H_k)_{\gamma_2,0}\!\! &\!\! (H_k)_{\gamma_2,1}\!\! &\!\! \cdots\!\! &\!\! (H_k)_{\gamma_2,\gamma_2}\!\!\!
\end{array}
\right], \label{eq:localizedmomentmatrix}
\end{equation}
where $\gamma_1$ and $\gamma_2$ are determined as follows.
\begin{equation}
\gamma_1=\left\{
\begin{array}{c}
(\mu+1)/2~~~({\rm if}~\mu~{\rm is~odd})\\
~~~~~\mu/2~~~~~~({\rm if}~\mu~{\rm is~even})
\end{array}
,\right.
\end{equation}
\begin{equation}
\gamma_2=\left\{
\begin{array}{c}
(\mu-1)/2~~~({\rm if}~\mu~{\rm is~odd})\\
~~~~~\mu/2~~~~~~({\rm if}~\mu~{\rm is~even})
\end{array}
.\right.
\end{equation}

Using $H^{(\gamma_1)}(\{m_{\bm{\beta}}\})$ and $H_k^{(\gamma_2)}(\{m_{\bm{\beta}}\})$, the following proposition provides linear matrix inequality (LMI) conditions that the moments of some (probability) measure on $\mathbb{R}_+^n$ must satisfy.\medskip\\
{\bf Proposition 1.}~{\rm \cite{Landau1987}}~Consider a sequence $(m_{\bm{\beta}})_{\bm{\beta} \in \mathcal{A}_{\mu+1}}$ with the set $\mathcal{A}_{\mu+1}$ defined by (\ref{eq:setAmu}). The sequence $(m_{\bm{\beta}})_{\bm{\beta}\in \mathcal{A}_{\mu+1}}$ constitutes moments of some measure defined on the positive orthant $\mathbb{R}_+^n\coloneqq\{\bm{x} \in \mathbb{R}^n~|~x_k \ge 0~(k=1,2,\cdots,n)\}$ only if 
\begin{align}
H^{(\gamma_1)}(\{m_{\bm{\beta}}\})&\succeq0,\label{eq:momentcondition}\\
H_k^{(\gamma_2)}(\{m_{\bm{\beta}}\})&\succeq0~~~(k=1,2,\cdots,n).\label{eq:localizedmomentcondition}
\end{align}

\medskip
\noindent
The conditions (\ref{eq:momentcondition}) and
(\ref{eq:localizedmomentcondition}) are called a moment condition and
localized moment conditions, respectively. Combining Proposition 1 with
(\ref{eq:momenteq0}),
the following theorem specifies semi-algebraic conditions that the moments of
the stochastic chemical reactions must satisfy.

\medskip
\noindent
{\bf Theorem 1.}~
Consider the stochastic chemical reaction modeled by the equation
(\ref{eq:CME}). Let $m_{\bm \beta}^*$ denote the steady state moments of
the random variable ${\bm x}$.
For a given truncation order $\mu$, the moments $m_{\bm \beta}^*$ satisfy the following conditions.
\begin{align}
&0=\sum_{\beta_1}\sum_{\beta_2}\cdots\sum_{\beta_n}A_{\bm{\alpha},\bm{\beta}}m_{\beta}^*~~~\left(\bm{\alpha}\in\mathcal{A}_{\mu}\subset\mathbb{N}_0^n\right)\nonumber,\\
&H^{(\gamma_1)}(\{m_{\beta}^*\})\succeq0, \label{eq:subject}\\
&H_k^{(\gamma_2)}(\{m_{\beta}^*\})\succeq0~~~(k=1,2,\cdots,n),\nonumber
\end{align}
where $A_{\bm{\alpha},\bm{\beta}}$ are the coefficients of the polynomial
(\ref{eq:momenteqcoefficient}).

\medskip
\noindent
Theorem 1 implies that the moments of the stochastic chemical reaction
governed by (\ref{eq:CME}) lie in the semi-algebraic set  (\ref{eq:subject}).
Thus, Problem 1 can be recast as a relaxation problem over the
 convex semi-algebraic set. In particular, we show, in the following section,
that the bounds of the mean and the variance can be computed by semidefinite
programming. 

\medskip
\noindent
{\bf Remark 1.}~
In the special case of $n=1$ random variable,
the conditions (\ref{eq:momentcondition}) and
(\ref{eq:localizedmomentcondition}) with $\gamma_1, \gamma_2
\rightarrow \infty$ become both necessary and sufficient for
the sequence $(m_\beta)_{\beta \in \mathbb{N}_0}$ to be the moments of
some measure on $\mathbb{R}_+$ \cite{Landau1987}.
The associated problem is called Stieltjes moment problem named after
the renowned mathematician Thomas Stieltjes. 
Similar conditions are available for the moments of measures defined on
various domains such as $[0,1]$, $\mathbb{R}$ and semi-algebraic sets
(see \cite{Landau1987, Shohat1943} for example).

\subsection{Semidefinite programming for mean and variance computation}
Theorem 1 asserts that the lower and the upper bounds of the steady state
statistics, say $f(m_{\bm{\alpha}})$, can be calculated by solving the following optimization problems, respectively.
\begin{equation}
\begin{array}{c}
~~~~~~~~~{\rm min}~~~~f(m_{\bm{\alpha}})\\
{\rm subject~to}~~~~~(\ref{eq:subject}),
\label{eq:optimizationmin}\end{array}
\end{equation}
\begin{equation}
\begin{array}{c}
~~~~~~~~~{\rm max}~~~~f(m_{\bm{\alpha}})\\
{\rm subject~to}~~~~~(\ref{eq:subject}).
\end{array}\label{eq:optimizationmax}
\end{equation}
Note that the constrains are convex. 
In what follows, we show that the bounds of  mean and variance, the two
most important statistics of stochastic chemical reactions, can be
 formulated as semidefinite programs.

 \medskip
 \noindent
{\bf Mean}~The objective function $f(m_{\bm{\alpha}})$ is obviously
 linear, and thus the problem falls into the class of semidefinite programming.
For example, if one wants to compute the mean of the copy number of the
 $i$-th molecule, the objective function should be
\begin{align}
 f(m_{\bm{\alpha}})= \mathbb{E}[x_i] = m_{\bm{e}_i}
\end{align}
with $\bm{e}_i$ representing the row vector whose $i$-th entry is 1 and
0 elsewhere.

\par
\smallskip
In general, the gap between the upper and the lower bounds decreases
monotonically with the increase of the truncation order $\mu$.
Thus, there is a tradeoff between computational cost and the
conservativeness of the bounds.

\medskip
\noindent
{\bf Upper bound of variance}~~~By definition, the variance of the copy number of the $i$-th molecule is given by 
\begin{align}
f(m_{\bm{\alpha}})&=\bm{e}_i~\mathbb{E}\left[(\bm{x}-\mathbb{E}[\bm{x}])(\bm{x}-\mathbb{E}[\bm{x}])^\mathrm{T}\right]\bm{e}_i^\mathrm{T}\nonumber\\
&=m_{2\bm{e}_i}-m_{\bm{e}_i}^2.\label{eq:nvarfm}
\end{align}
The second term of the objective function
$f(m_{\bm{\alpha}})=m_{2\bm{e}_i}-m_{\bm{e}_i}^2$ is quadratic, and the
optimization problem (\ref{eq:optimizationmax}) is seemingly not
semidefinite programming. However, using the Schur complement, we can
convert the optimization (\ref{eq:optimizationmax}) into an
equivalent semidefinite programming as follows.

\medskip
\noindent
{\bf Theorem 2.}~
The solution of the optimization problem (\ref{eq:optimizationmax}) with the objective function (\ref{eq:nvarfm}) is the same as that of the following optimization problem.
\begin{align}
{\rm max}~~~&~~~~~~~~~~~~V\nonumber\\
{\rm subject~to}~~~&~~~~~~~~~~(\ref{eq:subject})\label{eq:optimizationvar}\\
&\left[
\begin{array}{cc}
m_{2\bm{e}_i}-V & m_{\bm{e}_i}\\
m_{\bm{e}_i} & 1
\end{array}
\right]\succeq O.\nonumber
\end{align}\medskip\\
The proof of this theorem can be found in Appendix A. 

\medskip
\noindent
{\bf Lower bound of variance}
The lower bound of variance can be also computed by semidefinite
programming with additional relaxation. Specifically, we first
compute the upper bound of the mean value $m_{\bm{e}_i}$, and then substitute
the result into $m_{\bm{e}_i}^2$ of the equation (\ref{eq:nvarfm}) to
obtain a linear objective function.
The estimated lower bound of variance monotonically increases with
the truncation order $\mu$ since the upper bound of the mean value
$m_{\bm{e}_i}$ monotonically decreases.

\par
\smallskip
As the optimization problems shown in this subsection use convex relaxation, 
there may be the case where the truncation order $\mu$ needs to be
unrealistically large to reduce the conservativeness of the bounds to a practically useful level.
To the authors' experience, however, one will only need to use $\mu = 6$ to 12-th
order moments to obtain a practically useful bounds for small reaction
networks with  $n = 3$ to 5 molecules. The general quantitative
analysis, however, is still an open question.

\medskip
\noindent
{\bf Remark 2.}~The moment closure approaches
\cite{Zhao2010, Singh2011} allow for approximating the high order
moments, $m_{\bm{\beta}}$ with ${\rm deg}(m_{\bm{\beta}})=\mu+1$, using
the low order moments, $m_{\bm{\beta}}$ with
${\rm deg}(m_{\bm{\beta}})\le\mu$, to solve the equation
(\ref{eq:momenteq0}). This approach, however, is based on assumptions on
the underlying distribution $P_{\bm x}(t)$.
Thus, the quantification of the approximation error is not easy.
On the other hand, the proposed semidefinite programs are
 advantageous in that they can compute the upper and the lower bounds of the statistics of stochastic
 chemical reactions with theoretical rigor without assuming an underlying distribution.
 
\begin{figure}[tp]
\centering
\includegraphics[width=8.5cm]{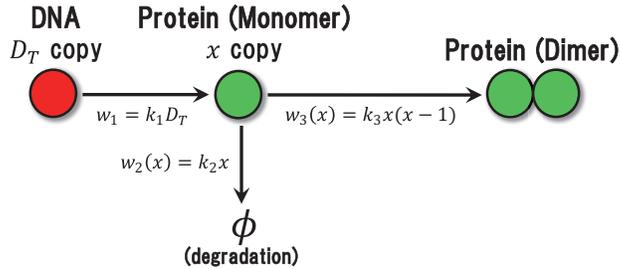}
\caption{Genetic circuit producing monomer and dimer proteins}
\label{fig:monomerdimerfig}
\end{figure}

\section{Application to a Stochastic Genetic Circuit}
To illustrate the proposed semidefinite programming
approach, we analyze a protein dimerization process,
which is one of the simplest examples of stochastic biomolecular reactions.
Consider the chemical reactions shown in Fig. \ref{fig:monomerdimerfig},
where there are three species of molecules, DNA, monomer protein
and dimer protein, and $r=3$ reactions.
In Fig. \ref{fig:monomerdimerfig}, the protein monomer is first produced
from DNA, and then it is either degraded or dimerized. As a result, the copy
number of protein monomer reaches a steady state at $t \rightarrow \infty$.

\par
\smallskip
In what follows, our goal is to analyze the steady state mean and variance of the
copy number of the protein monomer.
Let the copy number of monomer protein and its $\alpha$-th order moment
be defined by $x$ and 
\begin{equation}
m_{\alpha}\coloneqq\mathbb{E}\left[x^\alpha\right]=\sum_{x=0}^\infty
 x^\alpha P_x(t), \nonumber
\end{equation}
respectively. It should be noticed that $m_0=1$ by definition.
Let $D_T$ denote the total copy number of DNA molecules and assume that $D_T$ is a constant.
The reaction rates of monomer production, degradation and dimerization
$w_i({\bm x})~(i=1,2,3)$ are then defined by the polynomials in Table \ref{tbl:allwi}. 
The stoichiometry of the reactions, or the increment of the copy number of
monomer protein by each reaction, is given by
\begin{equation}
s_1=1,~~~s_2=-1,~~~s_3=-2.\nonumber
\end{equation}

\begin{table}[tp]
\begin{center}
\caption{Transition rates $w_i(x)$ in Fig.~\ref{fig:monomerdimerfig}}
\label{tbl:allwi}
\begin{tabular}{c|c}
\hline
Reaction $i$ & Transition rate $w_i(x)$\\
\hline
$1$&$k_1D_T$\\
$2$&$k_2x$\\
$3$&$k_3x(x-1)$\\
\hline
\end{tabular}
\end{center}
\end{table}

Suppose the values of the parameters are given by Table \ref{tbl:parameter}.
First, we derive the moment equation (\ref{eq:momenteq0}) by expanding
the polynomial (\ref{eq:momenteqcoefficient}).
Specifically, let the truncation order be $\mu=3$. Then, we have 
\begin{equation}
\bm{0}=\left[\begin{array}{ccccc}
0.0\!&\!0.0\!&\!0.0\!&\!0.0\!&\!0.0\\
35.0\!&\!0.2\!&\!-0.2\!&\!0.0\!&\!0.0\\
35.0\!&\!69.6\!&\!0.7\!&\!-0.4\!&\!0.0\\
35.0\!&\!105.8\!&\!105.5\!&\!1.7\!&\!-0.6
\end{array}\right]\!\!\left[
\begin{array}{c}
m_0\\
m_1\\
m_2\\
m_3\\
m_4
 \end{array}\right].
\label{example-matrix-eq}
\end{equation}
It should be noted that the low order moments $m_1, m_2$ and $m_3$
depend on the higher order moment $m_4$ since the transition rate
$w_3(x)$ is quadratic in $x$ (see Table \ref{tbl:allwi}).
The equation (\ref{example-matrix-eq}) implies that the moments of the
stochastic reactions are in the nullspace of the matrix. 
To further constrain the feasibility set, we use the moment condition
and the localized moment condition, which are given by
\begin{equation}
\left[\begin{array}{ccc}
m_0&m_1&m_2\\
m_1&m_2&m_3\\
m_2&m_3&m_4
\end{array}
\right]\succeq O,~~~\left[\begin{array}{cc}
m_1&m_2\\
m_2&m_3
\end{array}
\right]\succeq O,
\end{equation}
respectively.
In practice, the truncation order $\mu$ needs to be determined so that
the gap between the lower and the upper bounds is sufficiently
small. This might require computing the optimization problems
(\ref{eq:optimizationmin}) and (\ref{eq:optimizationmax})
multiple times. 

\begin{table}[tp]
\begin{center}
\caption{Parameters of chemical reactions in Fig.~\ref{fig:monomerdimerfig}}
\label{tbl:parameter}
\begin{tabular}{c|cc}
\hline
\!\!\!Parameter\!\!\!& Value & Meaning \\
\hline
$k_1$ & $\!0.7~{\rm min}^{-1}$\!\!&\!\!\!\!\!\!\!Transcription and translation rate\!\!\\
$k_2$ & $\!\!\!{\rm ln}(2)/20~{\rm copy}^{-1}\!\!\cdot\! {\rm min}^{-1}\!$ &\!\!\!\!Protein (monomer) degradation\!\!\\
$k_3$ & $0.1~{\rm copy}^{-1}\!\!\cdot\! {\rm min}^{-1}$ & Dimerization rate\\
$D_T$ & $50~{\rm copy}$ & Total copy of DNA\\
\hline
\end{tabular}
\end{center}
\end{table}
\begin{figure}[tp]
\centering
\includegraphics[width=9cm]{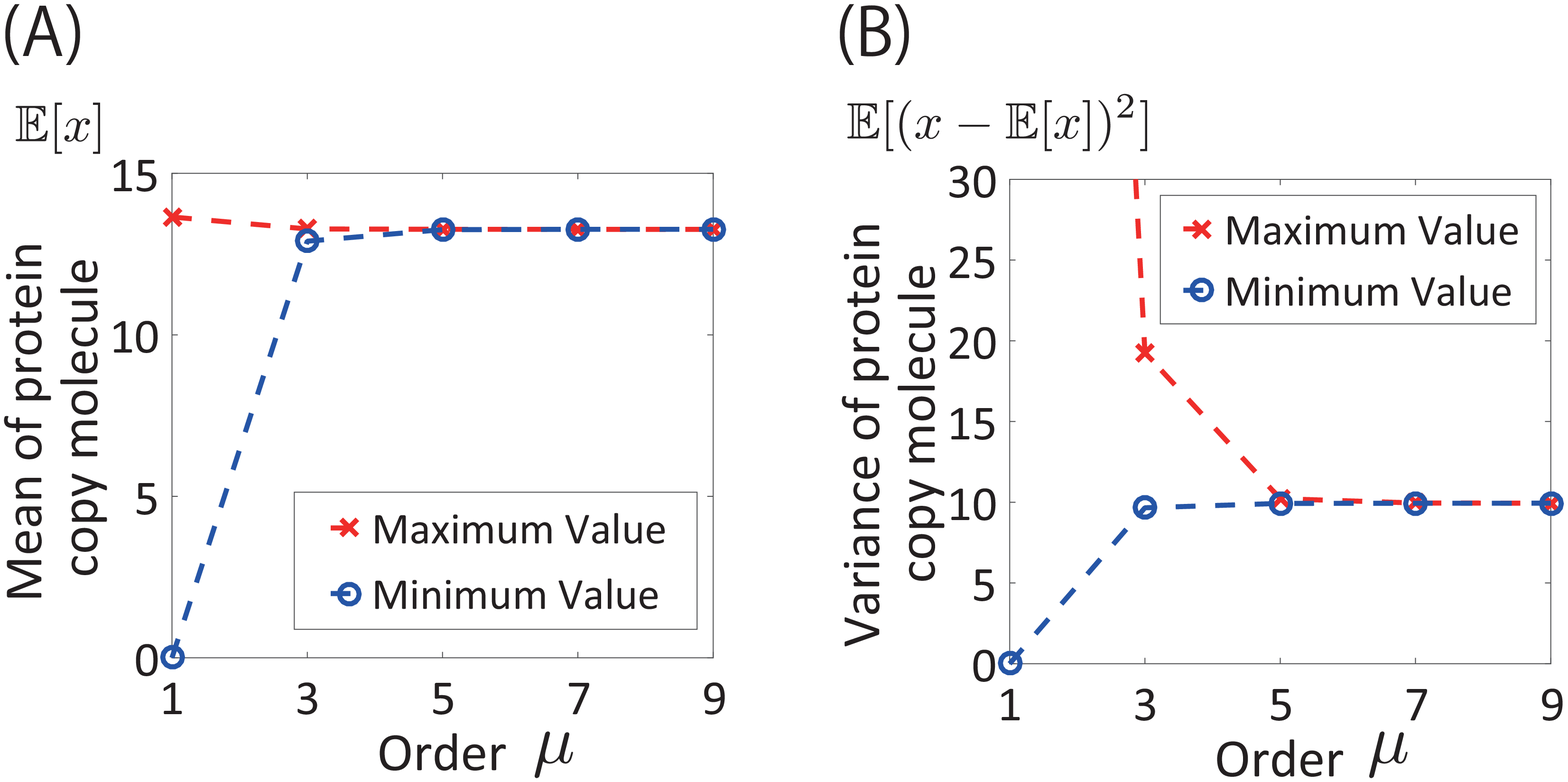}
 \caption{Upper and lower bounds of mean copy number and variance.}
\label{fig:AveVarvsOrder}
\end{figure}

\par
\smallskip
To analyze the mean copy number of protein monomer, let the objective
function be $f(m_{\bm{\alpha}})=\mathbb{E}[x]=m_{1}$ and solve the
optimization problems (\ref{eq:optimizationmin}) and
(\ref{eq:optimizationmax}).
Figure \ref{fig:AveVarvsOrder} (A) illustrates the computed lower and
the upper bounds of the mean monomer protein copy number for each truncation order $\mu$.
The gap between the bounds monotonically decreases with increasing
$\mu$, and with $\mu=5$, we can conclude 
 \begin{align}
	13.26 \le \mathbb{E}[x] \le 13.27, 
\end{align}
which is a sufficient resolution in practice.

\par
\smallskip
Next, to calculate the upper bound of the variance, let $f(m_{\bm \alpha}) =
\mathbb{E}[(x - \mathbb{E}[x])^2] = m_{2} - m_{1}^2$, and solve the
optimization problem  (\ref{eq:optimizationvar}).
The lower bound is also computed using the relaxation
described in Section III.
The optimization result in Fig. \ref{fig:AveVarvsOrder} (B) shows that the variance is bounded by
 \begin{align}
9.94 \le \mathbb{E}[(x - \mathbb{E}[x])^2] \le 9.95.
 \end{align}
It should be noticed that these results are mathematically strict and
thus enable rigorous quantification of biocircuit performance using the
statistical norms.

\medskip
\noindent
{\bf Remark 3.}~
The optimization problem was solved with SeDuMi 1.32 on MATLAB 2016b.
To avoid numerical issues, we normalized the variables $m_i$ by constants $20^i$ in the above numerical example.
It should be noted that optimization problems associated with the moment
matrices tend to be numerically unstable as reported in \cite{Kuntz2016}. 
\section{Conclusion}
In this paper, we have formulated the optimization problem to compute
the bounds of the steady state statistics of stochastic chemical reactions in genetic
circuits.
First, we have introduced the steady state moment equation, which is a
set of underdetermined linear equations.
To restrict the solution space of the moment equation, our key idea is to use
the LMI conditions, which the moments of some measure must satisfy.
Consequently, we have obtained a convex relaxation problem that can be solved
by semidefinite programming. 
A distinctive feature of the proposed approach is that it can provide mathematically rigorous upper and lower bounds of
the statistics for any stochastic chemical reactions modeled by the CME.
To demonstrate the method, we have analyzed the protein dimerization process and have obtained tight bounds of the mean and the variance.
 
 \par
 \smallskip
Although we have only considered the steady state moments, we can extend
 the proposed approach to the analysis of transient moments. The authors are currently
 working to build a stochastic biocircuit design tool based on the
 specifications of transient statistics.
\bibliographystyle{ieeetr}
\bibliography{arXiv_YS_YH}

\appendix
\section{Proof of Theorem 2.}
\noindent
{\bf Proof}~~~The optimization problem (\ref{eq:optimizationmax}) is equivalent to the following problem.
\begin{equation}
\begin{array}{c}
~~~{\rm max}~~~~~~~~~~~V\\
{\rm subject~to}~~~~f(m_{\bm{\alpha}})\ge V\\
~~~~~~~~~~~~~~~~~~(\ref{eq:subject}).
\end{array}
\notag
\end{equation}
As the objective function is $f(m_{\bm{\alpha}})=m_{2\bm{e}_i}-m_{\bm{e}_i}^2$, the inequality $f(m_{\bm{\alpha}})\ge V$ in the constraint can be equivalently written by
\begin{equation}
\left[
\begin{array}{cc}
m_{2\bm{e}_i}-V & m_{\bm{e}_i}\\
m_{\bm{e}_i} & 1
\end{array}
\right]\succeq O\label{eq:shurV}
\end{equation}
using Schur compliment \cite{Boyd2004}. The optimal value of
(\ref{eq:optimizationmax}) is thus equal to that of
(\ref{eq:optimizationvar}).
$\hfill \Box$

\addtolength{\textheight}{-3cm}   %
\end{document}